\documentclass[preprint,12pt]{elsarticle}  
\usepackage{graphicx} 
\usepackage{amssymb} 
\usepackage{amsmath} 

\journal{Physics Letters B} 

\begin{document} 

\begin{frontmatter} 

\title{Realistic calculations of ${\bar K}NN$, ${\bar K}NNN$, 
and ${\bar K}{\bar K}NN$ quasibound states} 

\author{N.~Barnea}
\author{A.~Gal\corref{cor1}} 
\cortext[cor1]{corresponding author: Avraham Gal, avragal@savion.huji.ac.il} 
\author{E.~Z.~Liverts}  
\address{Racah Institute of Physics, The Hebrew University, 91904 
Jerusalem, Israel} 
                     
\begin{abstract} 
Binding energies and widths of three-body $\bar KNN$, and of four-body 
$\bar KNNN$ and $\bar K\bar K NN$ nuclear quasibound states are calculated in 
the hyperspherical basis, using realistic $NN$ potentials and subthreshold 
energy dependent chiral ${\bar K}N$ interactions. Results of previous $K^-pp$ 
calculations are reproduced and an upper bound is placed on the binding energy 
of a $K^-d$ quasibound state. A self consistent handling 
of energy dependence is found to restrain binding, keeping the calculated 
four-body ground-state binding energies to relatively low values of about 30 
MeV. The lightest strangeness $-2$ particle-stable $\bar K$ nuclear cluster 
is most probably $\bar K\bar KNN$. The calculated $\bar KN\to\pi Y$ conversion 
widths range from approximately 30 MeV for the $\bar K NNN$ ground state to 
approximately 80 MeV for the ${\bar K}{\bar K}NN$ ground state. 
\end{abstract} 


\end{frontmatter}

\section{Introduction} 
\label{sec:intro} 

Unitarized coupled-channel chiral dynamics in the strangeness ${\cal S}=-1$ 
sector, constrained by fitting to $K^-p$ low-energy and threshold data, gives 
rise to a $(\bar K N)_{I=0}$ $s$-wave quasibound state (QBS), as detailed 
in recent works \cite{IHW11,CS12}. The relationship of this QBS to the 
observed $\Lambda(1405)$ resonance, which was predicted long ago by Dalitz 
and Tuan \cite{DT59} within a phenomenological study of $\bar K N-\pi\Sigma$ 
coupled channels, has been recently reviewed by Hyodo and Jido \cite{HJ12}. 
With that strong $(\bar K N)_{I=0}$ interaction, $\bar K$ mesons are expected 
to bind to nuclear clusters beginning with the $(\bar K NN)_{I=1/2}$ 
$J^{\pi}=0^-$ QBS, loosely termed $K^-pp$. While several few-body calculations 
confirmed that $K^-pp$ is bound, as reviewed in Ref.~\cite{weise10}, we here 
focus on those calculations using chiral interaction models in which the 
strong subthreshold energy dependence of the input $\bar K N$ interactions, 
essential in $\bar K$ nuclear few-body calculations, is under sound 
theoretical control. Such calculations yield binding energies in the range 
$B(K^-pp)\sim 10-20$ MeV \cite{DHW08,IKS10}, in contrast to values of 100 MeV 
or more obtained upon relegating peaks observed in final-state $\Lambda p$ 
invariant-mass spectra from FINUDA \cite{FINUDA05} and DISTO \cite{DISTO10} to 
the QBS decay $K^-pp\to\Lambda p$. To reinforce this discrepancy we note that 
none of the other published $K^-pp$ calculations based on $\bar K N$ 
phenomenology \cite{YA02,SGM07,IS07,WG09} managed to get as large $K^-pp$ 
binding energy as 100 MeV. 

Given this unsettled state of affairs for $K^-pp$, it is desirable to provide 
chiral model predictions for heavier $\bar K$ nuclear clusters starting with 
four-body systems and, in particular, to study the onset of binding for 
${\cal S}=-2$ clusters.{\footnote{We disregard the $\bar K\bar K N$ QBS which 
was calculated within a chiral interaction model to practically coincide with 
the $\bar K+(\bar K N)_{I=0}$ threshold \cite{KEJ08}.}} A good candidate is 
$\bar K\bar K NN$ which of all four-body $\bar K$ nuclear clusters has the 
largest number of $\bar K N$ bonds (four out of six). Furthermore, for the 
$I=0$, $J^{\pi}=0^+$ lowest energy QBS, and limiting the nuclear isospin 
to $I_N=1$ corresponding to the dominant $s$-wave $NN$ configuration, 
this QBS has the most advantageous $I_{\bar KN}=0,1$ composition of 
$V_{\bar KN}^{(I)}$, $3\div 1$ in favor of the strong $V_{\bar KN}^{(0)}$, 
same as in $K^-pp$. 

In this Letter we present fully four-body nonrelativistic calculations of 
the $\bar K$ nuclear clusters $\bar K NNN$ and ${\bar K}{\bar K}NN$ in the 
hyperspherical basis. Realistic $NN$ interactions and effective subthreshold 
$\bar K N$ interactions derived within a chiral model \cite{HW08} are used. 
The energy dependence of the subthreshold $\bar K N$ interactions is treated 
self consistently, extending a procedure suggested and practised in 
Refs.~\cite{CFGGM11,FG12,GM12}. This provides a robust mechanism to restrain 
the calculated binding energies of $\bar K$ nuclear clusters. Our calculations 
in the three-body sector reproduce the $K^-pp$ calculations of Dot\'{e} et 
al. \cite{DHW08} and provide an upper bound on the binding energy of a $K^-d$ 
$J^{\pi}=1^-$ QBS. In the four-body sector we find binding energies close 
to 30 MeV, in strong disagreement with predictions of over 100 MeV made 
in phenomenological, non-chiral models for $\bar KNNN$ \cite{AY02} and 
$\bar K\bar K NN$ \cite{YDA04,YAH11}.

\section{Input and Methodology} 
\label{sec:prelim} 

In this section we (i) briefly review the hyperspherical basis in 
which $\bar K$-nuclear cluster wavefunctions are expanded and in which 
calculations of ground-state energies are done, (ii) specify the two-body 
$NN,~\bar K N,~\bar K \bar K$ input interactions, and (iii) discuss the 
choice of $\bar K N$ subthreshold energy to be used self consistently in 
the binding energy calculations.

\subsection{Hyperspherical basis} 
\label{meth} 

The hyperspherical-harmonics (HH) formalism is used here similarly to its 
application in light nuclei \cite{BLO00} and recently in four-quark clusters 
\cite{VWBV07}. In the present case, the $N$--body wavefunction ($N=3,4$) 
consists of a sum over products of isospin, spin and spatial components, 
antisymmetrized with respect to nucleons and symmetrized with respect to 
$\bar K$ mesons. Focusing on the spatial components, translationally invariant 
basis functions are constructed in terms of one hyper-radial coordinate $\rho$ 
and a set of $3N-4$ angular coordinates [$\Omega_N$], substituting for $N-1$ 
Jacobi vectors. The spatial basis functions are of the form  
\begin{equation} 
\Phi_{n,[K]}(\rho,[\Omega_N])=R^{[N]}_{n}(\rho)
{\cal Y}^{[N]}_{[K]}([\Omega_N]), 
\label{eq:HH} 
\end{equation} 
where $R^{[N]}_{n}(\rho)$ are hyper-radial basis functions expressible in 
terms of Laguerre polynomials and ${\cal Y}^{[N]}_{[K]}([\Omega_N])$ are 
the HH functions in the angular coordinates $\Omega_N$ expressible in terms 
of spherical harmonics and Jacobi polynomials. Here, the symbol $[K]$ 
stands for a set of angular-momentum quantum numbers, including those 
of ${\hat L}^2$, ${\hat L}_z$ and ${\hat K}^2$, where ${\hat{\bf K}}$ 
is the total grand angular momentum which reduces to the total orbital 
angular momentum for $N=2$. The HH functions ${\cal Y}^{[N]}_{[K]}$ are 
eigenfunctions of ${\hat K}^2$ with eigenvalues $K(K+3N-5)$, and 
$\rho^{K}{\cal Y}^{[N]}_{[K]}$ are harmonic polynomials of degree $K$.

\subsection{Interactions} 
\label{sec:inter} 

For the $NN$ interaction we used the Argonne AV4' potential \cite{WP02} 
derived from the full AV18 potential by suppressing the spin-orbit and 
tensor interactions and readjusting the central spin and isospin dependent 
interactions. The AV4' potential provides an excellent approximation in 
$s$-shell nuclei to AV18. Its accuracy in $\bar K$ nuclear cluster 
calculations has been confirmed here by comparing our results for $K^-pp$ 
using AV4' with those of Ref.~\cite{DHW08} using AV18. 

For $\bar K h$ interactions, where the hadron $h$ is a nucleon or $\bar K$ 
meson, following Refs.~\cite{KEJ08,HW08} we have used a generic finite-range 
potential 
\begin{equation} 
V_{\bar K h}^{(I)}(r;\sqrt{s})=V_{\bar K h}^{(I)}(\sqrt{s})\exp(-r^2/b^2) 
\label{eq:Kbarh} 
\end{equation} 
with $b=0.47$ fm, where the superscript $I$ denotes the isospin of the 
$\bar K h$ pair and $\sqrt{s}$ is the Mandelstam variable reducing to the 
total energy in the two-body c.m. system. For $\bar K \bar K$, owing to 
Bose-Einstein statistics for $\bar K$ mesons, it is safe to assume that 
$V_{\bar K \bar K}^{(I=0)}=0$ at low energies where $s$ waves dominate. 
For $I_{\bar K \bar K}=1$, $V_{\bar K \bar K}^{(I=1)}=313$ MeV was obtained 
in Ref.~\cite{KEJ08} by fitting to the chiral leading-order Tomozawa-Weinberg 
$s$-wave scattering length. In the absence of nearby thresholds of coupled 
channels, no significant energy dependence is anticipated for this weakly 
repulsive $\bar K \bar K$ interaction. 

The $\bar K N$ interaction is an effective interaction based on chiral SU(3) 
meson-baryon coupled-channel dynamics with low-energy constants fitted to 
near-threshold $K^-p$ scattering and reaction data plus threshold branching 
ratios \cite{HW08}. Its HNJH version \cite{HNJH03} used here reproduces, 
a-posteriori, within error bars the $K^-p$ scattering length determined from 
the recent SIDDHARTA measurement of the $1s$ level shift and width of kaonic 
hydrogen \cite{SID11}. The {\it energy-dependent} complex potential strengths 
$V_{\bar K N}^{(I)}(\sqrt{s})$ were fitted in Ref.~\cite{HW08} by third-order 
polynomials in $\sqrt{s}$ in the range $1300\leq\sqrt{s}\leq 1450$ MeV, 
covering the $\pi\Sigma$ threshold at $\sqrt{s}\approx 1330$ MeV, as well 
as the $\bar K N$ threshold with isospin-averaged value $\sqrt{s_{\rm th}}=
1434.6$ MeV. The attractive real parts ${\rm Re}\,V_{\bar KN}^{(I)}(\sqrt{s})$ 
become gradually weaker for subthreshold arguments $\sqrt{s}\lesssim 1420$ 
MeV, a property shown below to be crucial in restraining the calculated 
binding energies of $\bar K$ nuclear clusters. The absorptive imaginary parts 
${\rm Im}\,V_{\bar K N}^{(I)}(\sqrt{s})$ that originate from $\bar KN\to\pi Y$ 
conversion also become weaker, but much faster, practically vanishing at the 
$\pi\Sigma$ threshold. 

\subsection{Energy dependence} 
\label{sec:sqrts}

The issue of energy dependence in near-threshold $\bar K N$ interactions 
deserves discussion. For a single $\bar K$ meson bound together with $A$ 
nucleons we define an average $\bar K N$ Mandelstam variable 
$\sqrt{s}_{\rm av}$ by 
\begin{equation}
A{\sqrt{s}}_{\rm av}={\sum_{i=1}^{A}\sqrt{(E_K+E_i)^2-({\vec p}_K+
{\vec p}_i)^2}}\;, 
\label{eq:sav}
\end{equation}
approximating it near threshold, $\sqrt{s_{\rm th}}\equiv m_N+m_K=1434.6$ 
MeV, by 
\begin{equation}
A{\sqrt{s}}_{\rm av}\approx A\sqrt{s_{\rm th}}-B-(A-1)B_K-{\sum_{i=1}^{A}
({\vec p}_K+{\vec p}_i)^2}/{2E_{\rm th}} \;,
\label{eq:thresh}
\end{equation} 
where $B$ is the total binding energy of the system and $B_K=-E_K$. 
Note that all the terms on the r.h.s. following $AE_{\rm th}$ are negative 
definite, so that ${\sqrt{s}}_{\rm av}\approx\sqrt{s_{\rm th}}+\delta\sqrt{s}$ 
with $\delta\sqrt{s}<0$. Hence, the relevant two-body energy argument of 
$V_{\bar K N}$ resides in the subthreshold region, forming a continuous 
distribution. The state of the art in non-Faddeev $\bar K$ nuclear 
calculations is to replace this distribution by an expectation value taken 
in the calculated QBS \cite{DHW08,CFGGM11,FG12,GM12}. Transforming squares 
of momenta in (\ref{eq:thresh}) to kinetic energies, the following expression 
is derived: 
\begin{equation} 
\langle\delta\sqrt{s}\rangle = -\frac{B}{A}-\frac{A-1}{A}B_K
-\xi_{N}\frac{A-1}{A}\langle T_{N:N} \rangle -\xi_{K}\left ( \frac{A-1}{A} 
\right )^2 \langle T_K \rangle \; , 
\label{eq:sqrt{s}} 
\end{equation} 
where $\xi_{N(K)}\equiv m_{N(K)}/(m_N+m_K)$, $T_K$ is the kaon kinetic 
energy operator in the total c.m. frame and $T_{N:N}$ is the pairwise $NN$ 
kinetic energy operator in the $NN$ pair c.m. system. Eq.~(\ref{eq:sqrt{s}}) 
refines the prescription $\langle\delta\sqrt{s}\rangle=-\eta B_K$, with 
$\eta=1,1/2$, used in the two types of $K^-pp$ variational calculations in 
Ref.~\cite{DHW08}. In the limit $A\gg 1$, it agrees with the nuclear-matter 
expression given in Ref.~\cite{CFGGM11} for use in kaonic atoms and $\bar K$ 
nuclear quasibound states. 

\begin{figure}[thb] 
\begin{center} 
\includegraphics[height=7cm]{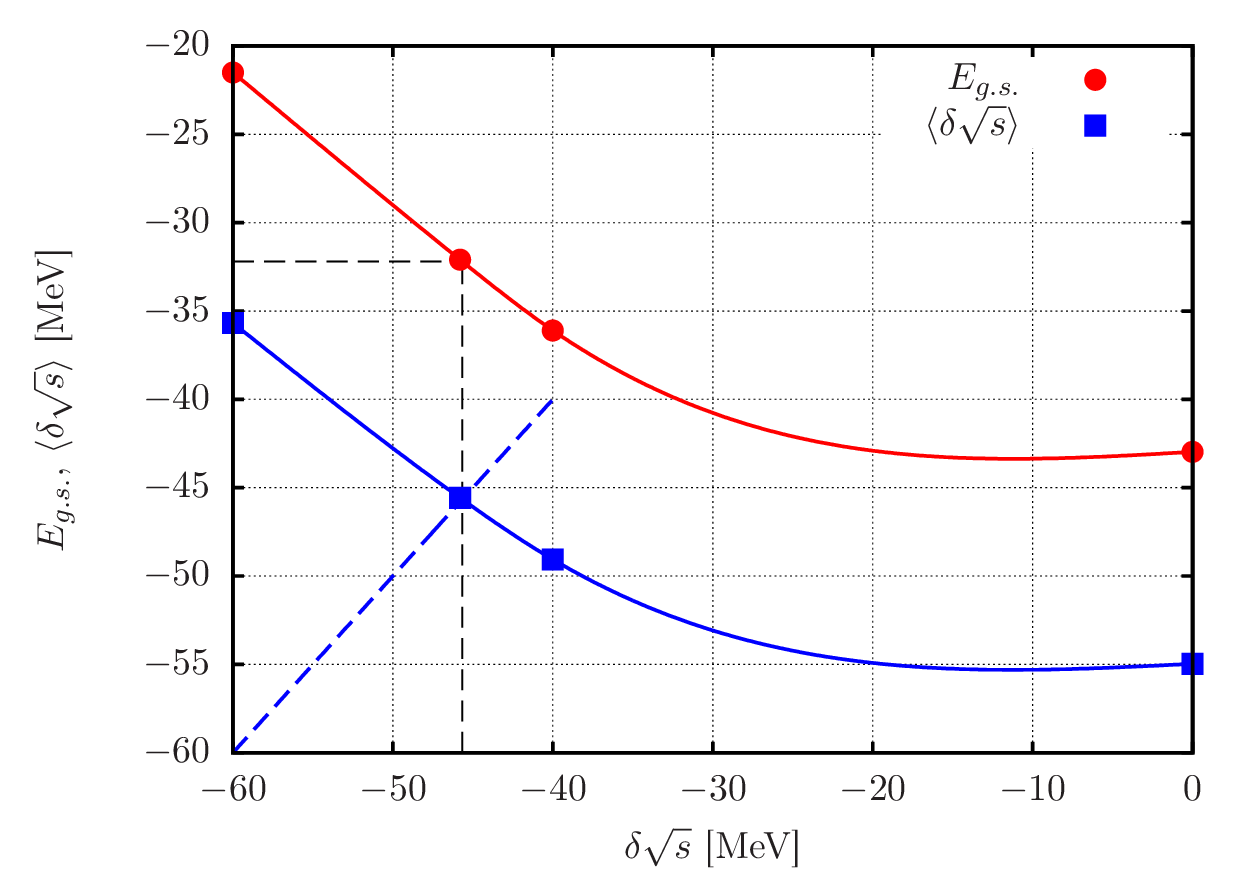} 
\caption{Self-consistency construction in $(\bar K\bar KNN)_{I=0,J^{\pi}=0^+}$ 
binding-energy calculations. The upper (red) and lower (blue) curves show 
calculated values of $E_{\rm g.s.}$ and $\langle\delta\sqrt{s}\rangle$ from 
Eq.~(\ref{eq:sqrt{s}NNKK}), respectively, vs. input $\delta\sqrt{s}$ values. 
The points connected by a vertical dashed line are the self-consistent values 
of $E_{\rm g.s.}$ and $\langle\delta\sqrt{s}\rangle$, the latter is obtained 
by the intersection of the dashed diagonal in the left-low corner with the 
lower (blue) curve.} 
\label{fig:bgl1} 
\end{center} 
\end{figure} 

A similar procedure is applied to the ${\bar K}{\bar K}NN$ system by summing 
up the four pairwise $\bar K N$ $\sqrt{s}$ contributions and expanding about 
$\sqrt{{s}_{\rm th}}$: 
\begin{equation}
\langle\delta\sqrt{s}\rangle=-\frac{1}{2}( B+\xi_{N}\langle T_{N:N} \rangle 
+\xi_{K}\langle T_{K:K} \rangle ) \; , 
\label{eq:sqrt{s}NNKK} 
\end{equation} 
where $T_{K:K}$ is the pairwise ${\bar K}{\bar K}$ kinetic energy operator 
in the ${\bar K}{\bar K}$ pair c.m. system. Eqs.~(\ref{eq:sqrt{s}}) and 
(\ref{eq:sqrt{s}NNKK}) provide a self-consistency cycle in $\bar K$ nuclear 
cluster calculations by requiring that the expectation value $\langle\delta
\sqrt{s}\rangle$ derived from the solution of the Schroedinger equation agrees 
with the input value $\delta\sqrt{s}$ used for $V_{\bar K N}(\sqrt{s})$. 
This is demonstrated in Fig.~\ref{fig:bgl1} for the lowest $\bar K\bar K NN$ 
configuration, with $I=0$, $J^{\pi}=0^+$. Its ground-state (g.s.) energy 
$E_{\rm g.s.}$, calculated upon suppressing ${\rm Im}\,V_{\bar K N}$, 
is shown by the upper (red) curve as a function of the input $\delta\sqrt{s}$. 
As one goes further down beginning approximately 15 MeV below threshold, 
the $\bar K N$ effective interaction from Ref.~\cite{HW08} becomes gradually 
weaker, resulting in less binding energy. In this subthreshold energy range 
the expectation values $\langle -\delta\sqrt{s}\rangle$, calculated from 
Eq.~(\ref{eq:sqrt{s}NNKK}) by solving the Schroedinger equation, also 
decrease upon increasing the input $-\delta\sqrt{s}$ values as shown by 
the lower (blue) curve. This allows to locate a self-consistent value 
$\langle\delta\sqrt{s}\rangle$ by drawing a diagonal to the lower curve 
and connecting it by a vertical line to the upper curve to identify the 
associated self-consistent value of $E_{\rm g.s.}$. 
For $({\bar K}{\bar K}NN)_{I=0,J^{\pi}=0^+}$, this construction yields 
a self-consistent value $\langle\delta\sqrt{s}\rangle=-46$ MeV and 
a corresponding value $E_{\rm g.s.}(\langle\delta\sqrt{s}\rangle)=-32.1$ MeV. 
If the energy dependence of $V_{\bar K N}(\sqrt{s})$ were neglected, 
and $V_{\bar K N}(\sqrt{s_{\rm th}})$ corresponding to $\delta\sqrt{s}=0$ 
were used instead, a considerably stronger binding would have emerged: 
$E_{\rm g.s.}(\delta\sqrt{s}=0)=-43.0$ MeV.

\section{Results and Discussion} 
\label{sec:res} 

We now present the results of self-consistent three-body and four-body 
calculations of $\bar K$ and $\bar K\bar K$ nuclear clusters. 
The three-body calculations have been tested by comparing with similar 
calculations for $K^-pp$ \cite{DHW08}. 

\begin{figure}[thb] 
\begin{center} 
\includegraphics[height=7cm]{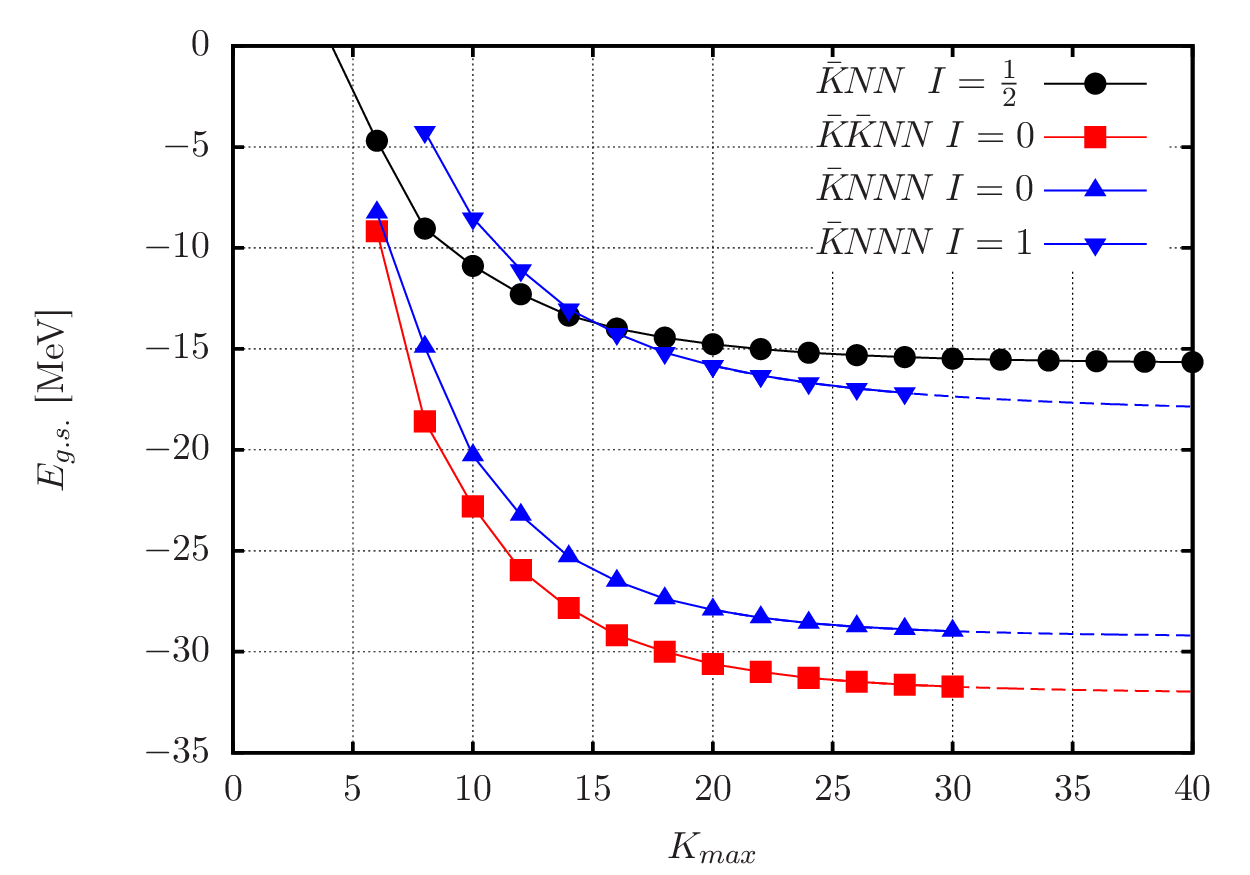}  
\caption{Ground-state energies of $\bar K$ nuclear clusters, calculated 
self consistently, as a function of $K_{\rm max}$. The dashed lines show 
extrapolation according to Eq.~(\ref{eq:conv}).} 
\label{fig:bgl2} 
\end{center} 
\end{figure} 

For a $\bar K$ nuclear cluster with global quantum numbers $I,L,S,J^{\pi}$, 
the potential and kinetic energy matrix elements were evaluated in the 
HH basis. The interactions specified in Section~\ref{sec:inter} conserve 
$L$ and $S$, the latter is given by the nuclear spin $S_N$. Since no 
$L\neq 0$ QBS are likely to become particle stable upon switching off 
${\rm Im}\,V_{\bar KN}$, we limit our considerations to $L=0$, resulting 
in $J=S=S_N$ with parity $\pm$ for even/odd number of $\bar K$ mesons, 
respectively. Although the total isospin $I$ is conserved by these 
charge-independent interactions, the isospin dependence of $V_{\bar K N}$ 
induces $\Delta I_N=1$ nuclear charge-exchange transitions, 
so that the nuclear isospin $I_N$ need not generally be conserved. 
Suppressing ${\rm Im}\,V_{\bar K N}$, the g.s. energy $E_{\rm g.s.}$ was 
calculated in a model space spanned by HH basis functions with eigenvalues 
$K\leq K_{\rm max}$. Self-consistent calculations were done for $\sqrt{s}$ 
from the $\bar KN$ threshold down to 80 MeV below, at which value the error 
incurred by the near-threshold approximation (\ref{eq:thresh}) is only 2.4 
MeV. Self consistency in $\delta\sqrt{s}$ was reached after typically five 
cycles. The convergence of binding energy calculations for particle-stable 
g.s. configurations is shown in Fig.~\ref{fig:bgl2} as a function of 
$K_{\rm max}$. With the exception of the $({\bar KNNN})_{I=1}$ cluster, 
good convergence was reached for values of $K_{\rm max}\approx 30-40$. 
The poorer convergence for $({\bar KNNN})_{I=1}$ is apparently due to its 
proximity to the $({\bar KNN})_{I=1/2}+N$ threshold. Asymptotic values of 
$E_{\rm g.s.}$ were found by fitting the constants $C$ and $\gamma$ of the 
parametrization 
\begin{equation} 
E(K_{\rm max})=E_{\rm g.s.}+\frac{C}{K_{\rm max}^{\gamma}} 
\label{eq:conv} 
\end{equation} 
to values of $E(K_{\rm max})$ calculated for sufficiently high values of 
$K_{\rm max}$. The accuracy reached is better than 0.1 MeV in the three-body 
calculations and about 0.2 MeV in the four-body calculations.  

\begin{figure}[thb]
\begin{center}
\includegraphics[height=7cm]{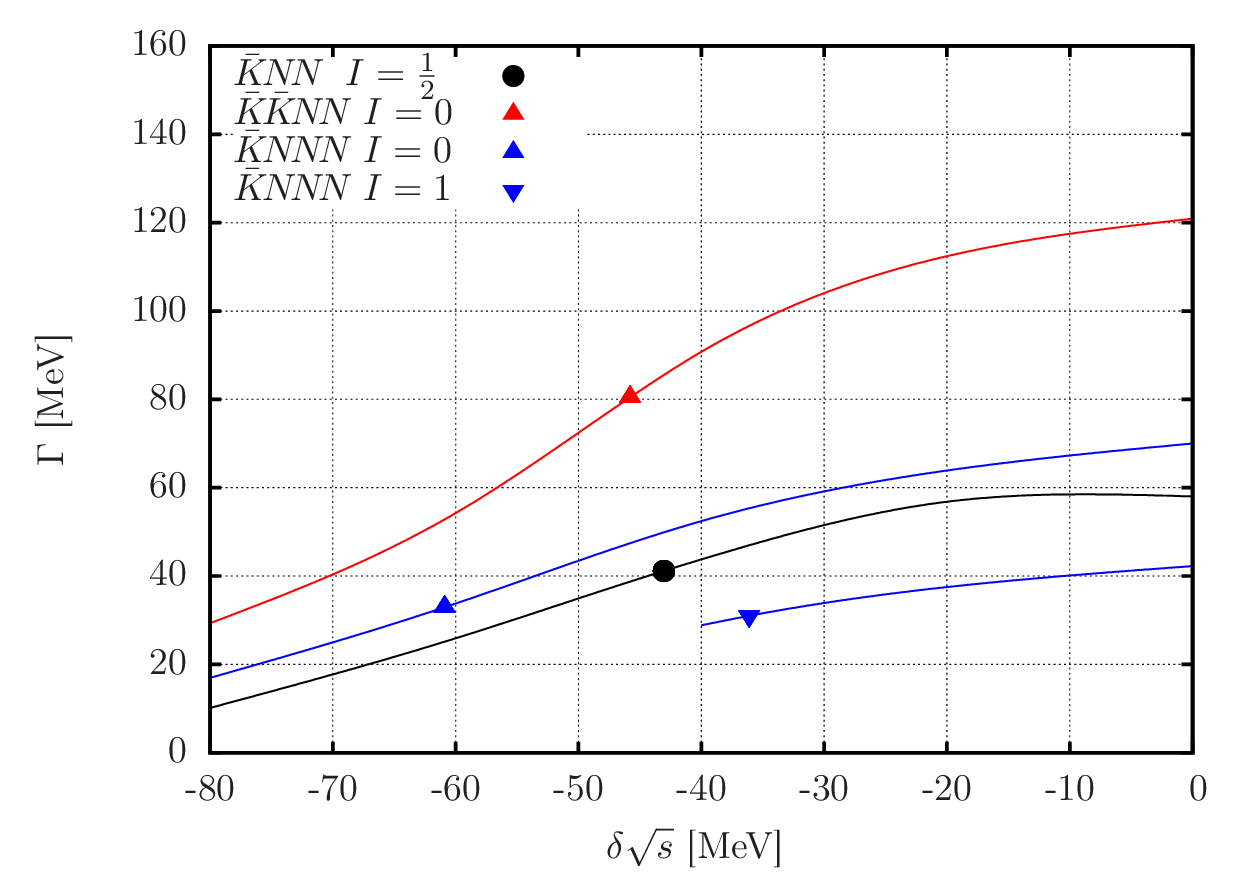} 
\caption{Conversion widths $\Gamma$ of $\bar K$ nuclear clusters calculated 
from Eq.~(\ref{eq:Gamma}) as a function of $\delta\sqrt{s}$. The widths 
appropriate to the self-consistent values $\langle\delta\sqrt{s}\rangle$ 
are denoted on each one of the curves.}
\label{fig:bgl3} 
\end{center} 
\end{figure} 

The conversion width $\Gamma$ was then evaluated through the expression 
\begin{equation}
\Gamma = -2\, \langle \,\Psi_{\rm g.s.}\, | \, {\rm Im}\,{\cal V}_{\bar KN} 
\, | \, \Psi_{\rm g.s.} \, \rangle \;,
\label{eq:Gamma} 
\end{equation} 
where ${\cal V}_{\bar KN}$ sums over all pairwise $\bar KN$ interactions. 
Since $|{\rm Im}\,{\cal V}_{\bar KN}|\ll |{\rm Re}\,{\cal V}_{\bar KN}|$, 
this is a reasonable approximation for the width. 
The dependence of the calculated width $\Gamma$ of $\bar K$ nuclear clusters 
on the input $\delta\sqrt{s}$ value used for the subthreshold $\bar K N$ 
energy is demonstrated in Fig.~\ref{fig:bgl3} for the same $\bar K$ nuclear 
clusters depicted in Fig.~\ref{fig:bgl2}. The width is seen almost invariably 
to decrease upon increasing $-\delta\sqrt{s}$, i.e. upon going deeper below 
threshold. This is similar to the dependence of  $E_{\rm g.s.}$ on the 
input $\delta\sqrt{s}$, as displayed for  $({\bar K}{\bar K}NN)_{I=0}$ 
in Fig.~\ref{fig:bgl1}. It is worth noting that the calculated widths of 
the single-$\bar K$ nuclear systems are clustered roughly in a range of 
$30-40$ MeV. Given a calculated width $\Gamma_{\bar KN}=43.6$ MeV for the 
underlying $(\bar K N)_{I=0}$ QBS, a scale of $\Gamma$(single $\bar K$) 
approximately 40 MeV appears quite natural. In contrast, the width calculated 
for the double-$\bar K$ system $({\bar K}{\bar K}NN)_{I=0}$ is about twice 
larger, approximately 80 MeV. 

In Table~\ref{tab:test} we compare results of the present work for 
$(\bar K N)_{I=0}$ and $(\bar K NN)_{I=1/2}$ QBS with those by Dot\'{e} et 
al.~\cite{DHW08}. Our $(\bar K N)_{I=0}$ calculation reproduces that of 
Ref.~\cite{KEJ08} and agrees with that in Ref.~\cite{DHW08} to within 0.1 
MeV out of binding energy $B\approx 11.5$ MeV and 0.2 MeV out of width 
$\Gamma\approx 43.7$ MeV, a precision of better than $1\%$. We note that this 
$\Lambda(1405)$-like QBS is bound considerably weaker than a QBS required by 
construction to reproduce $\Lambda(1405)$ nominally, with $B_{\Lambda(1405)}
\approx 27$ MeV \cite{AY02}. For a more complete discussion of this point 
we refer to \cite{HW08}. 

\begin{table}[htb] 
\begin{center} 
\caption{Comparison of $\bar KN$ and $\bar K NN$ QBS calculations, 
BGL (present) vs. DHW \cite{DHW08}.} 
\begin{tabular}{lcccccccc}  
\hline 
QBS & $I,J^{\pi}$ & Ref. & $\langle\delta\sqrt{s}\rangle$ & $B$ & $\Gamma$ & 
$B_K$ & $r_{NN}$ & $r_{KN}$ \\ 
 &  &  & [MeV] & [MeV] & [MeV] & [MeV] & [fm] & [fm] \\ 
\hline 
$\bar KN$ & $0,{\frac{1}{2}}^-$ &BGL& $-11.4$ & 11.4 & 43.6 & 11.4&--&1.87 \\ 
                &  & DHW & $-11.5$ & 11.5 & 43.8$^{\dag}$ & 11.5 &--& 1.86 \\ 
$\bar KNN$&$\frac{1}{2},0^-$&BGL& $-43$ & 15.7 & 41.2 & 35.5 & 2.41 & 2.15 \\ 
                &  & DHW & $-39$ & 16.9 & 47.0 & 38.9 & 2.21 & 1.97 \\ 
\& $I_N=1$   &  & BGL & $-35$ & 11.0 & 38.8 & 27.9 & 2.33 & 2.21 \\ 
                &  & DHW & $-31$ & 12.0 & 44.8 & 31.0 & 2.13 & 2.01 \\ 
\hline 
\end{tabular} 
\label{tab:test} 
$^{\dag}$we thank Dr. A.~Dot\'{e} for communicating to us this width value. 
\end{center} 
\end{table} 

For $\bar K NN$ with $I=1/2$ and $J^{\pi}=0^-$, loosely termed $K^-pp$, 
we compare the present calculation with the type-I HNJH-versed DHW 
variational calculation \cite{DHW08} for which the implied effective 
$\langle\delta\sqrt{s}\rangle$ value is close to our self-consistent 
$\langle\delta\sqrt{s}\rangle$ value. From their type-I,II calculations one 
concludes that $\delta B/\langle\delta\sqrt{s}\rangle\approx 0.24$, so that 
our binding energy value $B$ should come out smaller by approximately 1 MeV 
than their listed type-I $B$. The remainder 0.2 MeV of the 1.2 MeV difference 
between rows 3 and 4 in the table is attributed to using slightly different 
$NN$ interactions: AV4' here, AV18 in Ref.~\cite{DHW08}. Rows 5 and 6 of 
the table demonstrate the effect of limiting the model space to $I_N=1$, 
compatible with the dominant $s$-wave $NN$ configuration. This results in 
a decrease of the calculated binding energy by $4.8\pm 0.1$ MeV. The 1 MeV 
difference between rows 5 and 6 is consistent with the estimate made above for 
$\delta B/\langle\delta\sqrt{s}\rangle$, with no room within $NN$ $s$ waves 
for any marked difference arising from the difference between using AV4' (BGL) 
and AV18 (DHW). Finally, the differences of order $10-15\%$ between the two 
width calculations, and between the two r.m.s. distance calculations, reflect 
the sensitivity of these entities to details of the three-body wavefunction,  
particularly through the effective $\langle\delta\sqrt{s}\rangle$ value used. 

We have also searched for a $\bar K NN$ QBS with $I=1/2$ and $J^{\pi}=1^-$, 
loosely termed $K^-d$. The possibility of a QBS with these quantum 
numbers has hardly been discussed in the literature, apparently since 
it was realized from the very beginning \cite{nogami63} that $K^-d$ is 
less exposed than $K^-pp$, by a ratio close to $1\div 3$, to the strongly 
attractive $V_{\bar K N}^{(0)}$ interaction. We are not aware of any genuine 
three-body calculation for $K^-d$.{\footnote{However, very recently Oset 
et al. \cite{oset12} made an estimate within the Fixed Center Approximation 
for a $K^-d$ QBS with total binding energy $B=9$ MeV and conversion width 
$\Gamma\approx 30$ MeV. Alternatively, extrapolating below threshold the best 
educated guess for the scattering length $a_{K^-d}$ \cite{DM11} yields an 
estimate of $B=6.6$ MeV and $\Gamma\approx 29$ MeV.}} Our calculations did not 
produce any $I=1/2,J^{\pi}=1^-$ QBS below the $(\bar KN)_{I=0}+N$ threshold, 
i.e. with total binding energy exceeding 11 MeV. 
Whether or not such a QBS exists above the $(\bar KN)_{I=0}+N$ threshold is 
an open question which cannot be resolved within the present HH calculations 
that normally converge at the lowest energy state for given quantum numbers. 

In Table~\ref{tab:res} we present new results for $\bar KNNN$ and 
$\bar K\bar K NN$ QBS. The first two rows concern the $\bar KNNN$ system 
essentially based on the $I_N=1/2$ mirror nuclei $^3$H and $^3$He which 
are bound by 8.99 MeV in this calculation. The $\bar K$ nuclear interaction 
splits the two resultant $I=0,1$ $\bar KNNN$ QBS such that the $I=0$ QBS 
is the lower of the two. The 11 MeV isospin splitting is small compared 
to the approximately 30 MeV conversion width of each of these states. 
We note that the $I=0$ QBS is bound weakly compared to the tight binding 
over 100 MeV predicted for it by Akaishi and Yamazaki \cite{AY02}. Its 
spatial dimensions, with interparticle distances all exceeding 2 fm, also 
do not indicate a very tight structure. The imposition of self consistency 
in the binding energy calculation is responsible for the relatively low 
value $B(\bar KNNN)_{I=0}=29.3$ MeV, compared to a considerably higher 
value $B(\bar KNNN)_{I=0}^{\delta\sqrt{s}=0}=42.1$ MeV upon using the 
threshold $\bar KN$ interaction. Higher values $B(\bar KNNN)_{I=0,1}$ 
would also have been obtained had we used the self-consistency requirement 
$\langle\delta\sqrt{s}\rangle=-B_K$ \cite{DHW08} which for $K^-pp$ gave 
a value of $B$ close to the one found by using (\ref{eq:sqrt{s}}), see Table 
\ref{tab:test}.    

\begin{table}[htb] 
\begin{center} 
\caption{Results of $\bar KNNN$ and $\bar K\bar K NN$ QBS calculations.} 
\begin{tabular}{lcccccccc} 
\hline 
QBS & $I,J^{\pi}$ & $\langle\delta\sqrt{s}\rangle$ & $B$ & $\Gamma$ & $B_K$ & 
$r_{NN}$ & $r_{NK}$ & $r_{KK}$ \\ 
 &  & [MeV] & [MeV] & [MeV] & [MeV] & [fm] & [fm] & [fm] \\  
\hline 
$\bar KNNN$&$0,{\frac{1}{2}}^+$&$-61$& 29.3 & 32.9 & 36.6 & 2.07 & 2.05 &-- \\ 
       & $1,{\frac{1}{2}}^+$ & $-36$ & 18.5 & 31.0 & 21.0 & 2.33 & 2.55 &-- \\ 
$\bar K\bar KNN$ & $0,0^+$ & $-46$ & 32.1 & 80.5 & 33.6 & 1.84 & 1.88 & 2.31\\ 
\& $V_{\bar K\bar K}=0$ &  & $-52$ & 36.1 & 83.2 & 37.9 & 1.71 & 1.70 & 2.01\\ 
\hline 
\end{tabular} 
\label{tab:res} 
\end{center} 
\end{table} 

The last two rows of Table~\ref{tab:res} report on the ${\cal S}=-2$ 
$(\bar K\bar K NN)_{I=0}$ QBS which has been highlighted as a possible 
gateway to kaon condensation in self-bound systems, given its large 
binding energy over 100 MeV predicted by Yamazaki et al. \cite{YDA04}. Our 
calculated value $B=32.1$ MeV is comparable with that for the ${\cal S}=-1$ 
$(\bar K NNN)_{I=0}$ QBS, and is a factor of two larger than for the lowest 
$\bar K NN$ QBS with $I=1/2$ and $J^{\pi}=0^-$. Note, however, that 
$(\bar K\bar K NN)_{I=0}$ is bound by less than 10 MeV with respect 
to the threshold for decay to a pair of $(\bar K N)_{I=0}$ 
$\Lambda(1405)$-like QBS. This apparent relatively weak binding of 
$(\bar K\bar K NN)_{I=0}$ is owing to the restraining effect of 
handling self consistently the energy dependent $\bar K N$ interaction.
Finally, the last row of the table shows what happens when the repulsive 
$V_{\bar K\bar K}$ is switched off. The effect is mild, increasing $B$ 
by only 4 MeV. Nevertheless, inspection of the r.m.s. distances in 
$(\bar K\bar K NN)_{I=0}$ reveals a more compact structure than 
$(\bar K NNN)_{I=0}$, which is also reflected by the large value of 
$\Gamma(\bar K\bar K NN)_{I=0}$.

\section{Conclusion} 
\label{sec:concl} 

In conclusion, we have performed calculations of three-body $\bar K NN$ and 
four-body $\bar KNNN$ and $\bar K\bar KNN$ QBS systems. Using practically 
identical interactions to those used in the $K^-pp$ chiral model calculations 
by Dot\'{e} et al. \cite{DHW08}, we were able to test our calculations for 
this QBS against theirs. Given the low binding energy $B(K^-pp)\approx 16$ MeV 
and sizable conversion width $\Gamma_{\rm conv}(K^-pp)\approx 40$ MeV, 
it might be difficult to identify such a near-threshold QBS unambiguously in 
ongoing experimental searches. This situation gets further complicated by two 
additional factors: (i) the possible presence of a near-threshold $K^-d$ QBS 
in the same charge state as the one in which $K^-pp$ is searched on, and (ii) 
additional two-nucleon absorption widths $\Delta\Gamma_{\rm abs}$ accounting 
for the poorly understood non-pionic processes $\bar K NN\to YN$. 
For $K^-pp$ we note the estimate $\Delta\Gamma_{\rm abs}(K^-pp)\lesssim 10$ 
MeV \cite{DHW08}. Appreciable $p$-wave contributions to the $K^-pp$ width were 
also suggested in Ref.~\cite{DHW08}, but doubts have been recently expressed 
on the effectiveness of a $p$-wave $\bar KN$ interaction by testing its role 
in kaonic atoms \cite{CFGGM11}. Altogether, the widths of $\bar K NN$ QBS are 
likely to be dominated by their conversion widths. 

For the four-body QBS systems ${\bar K}NNN$ and ${\bar K}{\bar K}NN$ we found 
relatively modest binding, of order 30 MeV in both, with conversion widths 
ranging from about 30 MeV for each of the ${\bar K}NNN$ QBS to about 80 MeV 
for the lowest ${\bar K}{\bar K}NN$ QBS. These systems, although somewhat 
more compact than $K^-pp$, are not as compact as suggested by Yamazaki et 
al. \cite{AY02,YDA04,YAH11}. Their $\bar K N$ r.m.s. distances do not fall 
below that of the $\Lambda(1405)$-like $\bar K N$ QBS, and their $NN$ r.m.s. 
distances exceed that of nuclear matter ($\approx 1.7$ fm). For a conservative 
estimate of the absorption widths $\Delta\Gamma_{\rm abs}$ in these systems, 
we count the number of nucleons $n$ available to join a given $\bar K N$ 
correlated pair, one pair per each $\bar K$. This gives twice as large $n$ 
for each of the four-body systems ($n=2$) with respect to $K^-pp$ ($n=1$). 
Hence, neglecting three-nucleon absorption, 
$\Delta\Gamma_{\rm abs}({\bar K}NNN,{\bar K}{\bar K}NN)\sim 20$ MeV. 

The energy dependence of the subthreshold $\bar K N$ effective interaction, 
constructed in Ref.~\cite{HW08} within a coupled channel chiral model, was 
found to be instrumental in restraining the binding of the four-body 
$\bar K$ nuclear clusters through the self-consistency requirement derived 
here for these light systems. A strong $\bar K N$ interaction operates to form 
tightly bound compact structures, necessarily accompanied by large kinetic 
energies. This leads by Eqs.~(\ref{eq:sqrt{s}}) and (\ref{eq:sqrt{s}NNKK}) to 
substantial values of the energy shift $\langle\delta\sqrt{s}\rangle$ which 
give rise to weaker input $\bar K N$ interactions, resulting in less binding 
as demonstrated in Fig.~\ref{fig:bgl1} for ${\bar K}{\bar K}NN$. However, 
dispersive contributions to the binding energy of QBS cannot be excluded. 
Recent fits to kaonic atoms \cite{CFGGM11,FG12} suggest that 
$\Delta B_{\rm disp}\sim\Delta\Gamma_{\rm abs}$, so that these 
binding energies could reach values $B(K^-pp)\sim 25$ MeV and 
$B(\bar K NNN,\bar K\bar K NN)\sim 50$ MeV. For heavier $\bar K$-nuclear 
clusters where the nuclear density is closer to nuclear-matter density, 
a restraining mechanism similar to the one discussed here has been shown 
to be operative \cite{GM12}. Other restraining, or saturation mechanisms are 
likely to be operative such as the increased ${\bar K}{\bar K}$ repulsion upon 
adding $\bar K$ mesons \cite{GFGM07}. It is therefore quite unlikely that 
strange dense matter is realized through $\bar K$ mesons as argued repeatedly 
by Yamazaki et al. \cite{YDA04,YAH11}.

\section*{Acknowledgements} 
This work was supported in part (NB and EZL) by the Israel Science Foundation 
grant 954/09, and in part (AG) by the EU initiative FP7, HadronPhysics2, under 
Project No. 227431.


\begin{thebibliography}{99} 


\bibitem{IHW11} Y.~Ikeda, T.~Hyodo, W.~Weise, Phys. Lett. B 706 (2011) 63, 
Nucl. Phys. A 881 (2012) 98. 

\bibitem{CS12} A.~Ciepl\'{y}, J.~Smejkal, Nucl. Phys. A 881 (2012) 115. 

\bibitem{DT59} R.H.~Dalitz, S.F.~Tuan, Phys. Rev. Lett. 2 (1959) 425, 
Ann. Phys. 10 (1960) 307. 

\bibitem{HJ12} T.~Hyodo, D.~Jido, Prog. Part. Nucl. Phys. 67 (2012) 55, 
and references therein.

\bibitem{weise10} W.~Weise, Nucl. Phys. A 835 (2010) 51, and references 
therein. 

\bibitem{DHW08} A.~Dot\'{e}, T.~Hyodo, W.~Weise, Nucl. Phys. A 804 (2008) 197, 
Phys. Rev. C 79 (2009) 014003. 

\bibitem{IKS10} Y.~Ikeda, H.~Kamano, T.~Sato, Prog. Theor. Phys. 124 (2010) 
533. 

\bibitem{FINUDA05} M.~Agnello, et al. [FINUDA Collaboration], Phys. Rev. Lett. 
94 (2005) 212303. 

\bibitem{DISTO10} T.~Yamazaki, et al. [DISTO experiment], Phys. Rev. Lett. 
104 (2010) 132502. 

\bibitem{YA02} T.~Yamazaki, Y.~Akaishi, Phys. Lett. B 535 (2002) 70. 

\bibitem{SGM07} N.V.~Shevchenko, A.~Gal, J.~Mare\^{s}, Phys. Rev. Lett. 98 
(2007) 082301; N.V.~Shevchenko, A.~Gal, J.~Mare\^{s}, J.~R\'{e}vai, 
Phys. Rev. C 76 (2007) 044004. 

\bibitem{IS07} Y.~Ikeda, T.~Sato, Phys. Rev. C 76 (2007) 035203; ibid. 79 
(2009) 035201. 

\bibitem{WG09} S.~Wycech, A.M.~Green, Phys. Rev. C 79 (2009) 014001. 

\bibitem{KEJ08} Y.~Kanada-E\'{n}yo, D.~Jido, Phys. Rev. C 78 (2008) 025212. 

\bibitem{HW08} T.~Hyodo, W.~Weise, Phys. Rev. C 77 (2008) 035204. 

\bibitem{CFGGM11} A.~Ciepl\'{y}, E.~Friedman, A.~Gal, D.~Gazda, J.~Mare\v{s},
Phys. Lett. B 702 (2011) 402, Phys. Rev. C 84 (2011) 045206. 

\bibitem{FG12} E.~Friedman, A.~Gal, Nucl. Phys. A 881 (2012) 150. 

\bibitem{GM12} D.~Gazda, J.~Mare\v{s}, Nucl. Phys. A 881 (2012) 159. 

\bibitem{AY02} Y.~Akaishi, T.~Yamazaki, Phys. Rev. C 65 (2002) 044005. 

\bibitem{YDA04} T.~Yamazaki, A.~Dot\'{e}, Y.~Akaishi, Phys. Lett. B 587 (2004) 
167. 

\bibitem{YAH11} T.~Yamazaki, Y.~Akaishi, M.~Hassanvand, Proc. Jpn. Acad. 
Ser. B 87 (2011) 362; M.~Hassanvand, Y.~Akaishi, T.~Yamazaki, Phys. Rev. C 84 
(2011) 015207. 

\bibitem{BLO00} N.~Barnea, W.~Leidemann, G.~Orlandini, Phys. Rev. C 61 (2000) 
054001. 

\bibitem{VWBV07} J.~Vijande, E.~Weissman, N.~Barnea, A.~Valcarce, Phys. Rev. D 
76 (2007) 094022. 

\bibitem{WP02} R.B.~Wiringa, S.C.~Pieper, Phys. Rev. Lett. 89 (2002) 182501. 

\bibitem{HNJH03} T.~Hyodo, S.I.~Nam, D.~Jido, A.~Hosaka, Phys. Rev. C 68 
(2003) 018201. 

\bibitem{SID11} M.~Bazzi, et al. [SIDDHARTA Collaboration], Phys. Lett. B 704 
(2011) 113, Nucl. Phys. A 881 (2012) 88. 

\bibitem{nogami63} Y.~Nogami, Phys. Lett. 7 (1963) 288. 

\bibitem{oset12} E.~Oset, D.~Jido, T.~Sekihara, A.~Martinez~Torres, 
K.P.~Khemchandani, M.~Bayar, J.~Yamagata-Sekihara, Nucl. Phys. A 881 (2012) 
127, and references therein.  

\bibitem{DM11} M.~D\"{o}ring, U.-G.~Mei{\ss}ner, Phys. Lett. B 704 (2011) 663. 

\bibitem{GFGM07} D.~Gazda, E.~Friedman, A.~Gal, J.~Mare\v{s}, Phys. Rev. C 76 
(2007) 055204; ibid. 77 (2008) 045206; ibid. 80 (2009) 035205. 





\end{thebibliography}
\end{document}